\documentstyle[prd,aps]{revtex}
\newcommand{\be}{\begin{equation}}
\newcommand{\ee}{\end{equation}}
\newcommand{\ba}{\begin{eqnarray}}
\newcommand{\ea}{\end{eqnarray}}
\newcommand{\oh}{\displaystyle{\frac{1}{2}}}
\begin{document}
\draft
%\twocolumn
%\preprint{MIT-CTP 2659, La Plata-Th 97/17, hep-th/9707199}
\title{Abelian and Non-Abelian Induced Parity Breaking Terms at Finite
Temperature}
\author{C.D.~Fosco$^a$\thanks{CONICET}\,,
G.L.~Rossini$^{b,c}$\thanks{CONICET.
On leave from La Plata University,
Argentina}\,
and\,
F.A.~Schaposnik$^c$\thanks{Investigador CICBA, Argentina}
\\
{\normalsize\it
$^a$Centro At\'omico Bariloche,
8400 Bariloche, Argentina}\\
{\normalsize\it
$^b$ Center for Theoretical Physics, Laboratory for
Nuclear Science and Department of Physics}\\
{\normalsize\it Massachusetts Institute of Technology,
 Cambridge, Massachusetts 02139, USA
}\\
{\normalsize\it
$^c$Departamento de F\'\i sica, Universidad Nacional de La Plata}\\
{\normalsize\it
C.C. 67, 1900 La Plata, Argentina}\\
(MIT-CTP-2659 ~ La Plata-Th 97/17 ~ hep-th/9707199)}

\date{ }
\maketitle
%===================================================================
\begin{abstract}
We compute the exact canonically induced parity breaking part of the 
effective action for 2+1 massive fermions in particular Abelian and 
non Abelian gauge field backgrounds. The method of computation resorts
to the chiral anomaly of the dimensionally reduced theory.
\end{abstract}

\bigskip

%===================================================================
%\newpage
\section{Introduction and results}

Three dimensional gauge theories coupled to matter are relevant
both in Field Theory and Condensed Matter physics. An important
feature of these theories is that, appart from the usual Maxwell
or Yang-Mills actions, there exists the possibility of
considering a Chern-Simons (CS) term as a part  \cite{djt}
or as the entire \cite{wi} gauge
field action. Moreover, even if the CS term is not included 
{\it ab initio}, it will be induced through fluctuation of Fermi
fields \cite{det}-\cite{NS0}, by the parity violating fermion mass
and/or by the celebrated parity anomaly \cite{djt}.

A fundamental property of the CS action is that its presence
forces a quantization law: the (non-Abelian) CS term is non-invariant
under ``large'' gauge transformations (i.e. gauge transformations
carrying non-trivial winding number), this implying that the coefficient
of the CS term should be quantized so that $\exp(iS_{CS})$ remains
single valued. Concerning the induced (through matter fluctuations)
CS term, it is well established that any gauge invariant regularization
of the massless fermionic determinant introduces 
a parity anomaly in the form of
$\pm\frac{1}{2} S_{CS}$ whose gauge non-invariance compensates the gauge 
non-invariance of the otherwise parity preserving effective action
\cite{det}. This parity anomalous contribution is also present in the
case of massive fermions \cite{GRS}, when other canonical parity violating 
terms associated to the fermion mass come into play.

The results above correspond to Quantum Field Theory at zero temperature.
What about $T \ne 0$ ? To our knowledge this question was first addressed
in \cite{pis} where it was argued that the coefficient of the induced
CS term remains unchanged at finite temperature. Contrasting with this
analysis, perturbative calculations 
yielded effective actions with CS coefficients which are smooth
functions of the temperature \cite{NS}-\cite{I}. It is important to notice 
that these computations dealt with the fermion mass dependent parity breaking
and ignored the parity anomaly related to gauge invariant regularizations.

The issue of renormalization of the CS coefficient induced by fermions
at $T \ne 0$ was reanalysed in refs.\cite{bfs}-\cite{cfrs} where it was
concluded that, in perturbation theory and on gauge invariance grounds, the
effective action for the gauge field cannot contain the smoothly 
renormalized CS coefficient which was the answer of perturbative 
calculations. More recently, the exact result for the effective action of a
$0+1$ massive fermion system \cite{dll} as well as 
non-perturbative calculations of the effective action in the $2+1$ Abelian 
case \cite{dgs} and its explicit
temperature dependent parity breaking part \cite{frs} 
have explicitly shown that although the 
perturbative expansion leads to a non-quantized $T$-dependent 
CS coefficient, the complete effective action can be seen to
be gauge invariant under both small and large gauge transformations,
the temperature depending shift in the CS coefficient being
just a byproduct of considering just the first term in the expansion
of the effective action.
% aca miento un poco; ya aprendi que con la parte que viola paridad 
% sola no se puede discutir invarianza de gauge.

We extend in the present work the analysis presented in \cite{frs}
for the Abelian model to the case of $2+1$ massive fermions in a 
non-Abelian gauge background.
By considering a particular class of gauge field background configurations
we compute exactly the induced parity breaking part of the
effective action for three dimensional massive fermions in the
fundamental representation of $SU(N)$.

To be precise, we are concerned with
\be
\Gamma_{odd} (A,M)  \;=\;\oh (\Gamma (A,M)-\Gamma (A,-M)) 
\label{gammaodd}
\ee
where  
\be
\exp\left( {- \Gamma (A,M)} \right)\;=\;  \int {\cal D} \psi \,
{\cal D} {\bar \psi} \; \exp \left[ - S_F (A, M) \right] ,
\label{ealagamma}
\ee
and
$S_F (A, M)$ is the  action for
massive fermions (with mass $M$) in a gauge background
$A_{\mu}$. 
As mentioned  above (see \cite{dgs} for a discussion), 
the mass dependent parity violating term is not the only 
one arising in $\Gamma(A,M)$;
there is also a local parity anomaly contribution in the form of half 
a CS term
arising in any gauge invariant regularization. This term, first noticed at 
$T=0$ in \cite{det} for massless fermions and in \cite{GRS} for massive 
fermions, is mass and temperature independent and can be removed by a local 
counterterm at the price of breaking large gauge invariance. It is in fact
not taken into account  
in most of the literature analysing $2+1$ dimensional massive fermion
models. To understand the interplay between the two contributions,
one can regard
the mass dependent parity violating term as naturally arising 
due to the fact that the Lagrangian already
contains at the classical level
a parity violating mass term. Concerning the mass-independent 
contribution which comes from the
parity anomaly, it can be seen as a necessary consequence of 
any gauge invariant regularization of the path integral fermionic measure.
After these remarks, it is clear that our definition of $\Gamma_{odd}$
excludes this last anomalous contribution but since
it is temperature independent, it does not affect
our analysis.

The calculation of (\ref{gammaodd}) for the general case, namely, for
{\em any \/} gauge field configuration is not something we can
do exactly. Instead of making a perturbative calculation dealing
with a small but otherwise arbitrary gauge field configuration,
we shall consider a restricted set of gauge field configurations
which can however be treated exactly. 

In order to get an exact result we 
choose a particular gauge field background which corresponds to 
a vanishing color electric field and a time-independent color magnetic field,
\be
A_3 = A_3(\tau), 
\label{0}
\ee
\be
 A_j = A_j(x)\,\,\, (j=1,2) 
\label{1}
\ee
or any equivalent configuration by gauge transformations.
In the non-Abelian case we further restrict $A_3$ to point in 
a fixed direction in the internal space,
\be
A_3 = |A_3| \check{n}, 
\label{a3}
\ee
and $A_j$ to commute with $A_3$,
\be
 [A_j , A_3]=0
\,\,\, (j=1,2) \;.
\label{aj}
\ee
Although for $SU(2)$ this implies that all of the components of $A_{\mu}$ 
commute, and can be thus seen as an ``Abelian-like'' configuration,
for $SU(N)$ with $N>2$ one can see that genuine  non-Abelian effects
are incorporated. The configurations under consideration are reminiscent of the
ones treated in \cite{NS0} for massless fermions at $T=0$; in that case Lorentz
covariance of the local result allowed straightforward generalization to 
arbitrary backgrounds. Unfortunately, this will not be the case here.

Our main result can be presented through the formula we obtain for 
$\Gamma_{odd}$,
\be
\Gamma_{odd} \;=\;
\frac{ig}{4\pi } tr \left(\arctan[\tanh(\frac{\beta M}{2}) \tan( 
\frac{g}{2} \int_0^{\beta}A_3 d\tau)]
\int d^2x \, \epsilon_{ij} F_{ij}\right)
\label{2}
\ee
where $g$ is the  coupling constant, $\beta = 1/T$ and $tr$ is an adequate
trace in $SU(N)$ 
(matrix functions are defined as usual as power series).

The paper is organized as follows. We give in section II a more
detailed description of the results presented in \cite{frs}
for the Abelian case so as to clarify the method of computation, which 
relies on the ability to factorize the piece of the efective action 
depending on the sign of the fermion mass.
The same method is
applied in section III to the analysis of the $SU(N)$ case
leading to formula (\ref{2}). Finally, in section IV
we summarize and give a discussion of our results.

\section{The Abelian case}

We are interested in evaluating the parity-odd piece
of the effective action (\ref{gammaodd}) 
which is induced by integrating out massive fermions coupled
to an Abelian gauge field $A_\mu$ in $2+1$ dimensions at finite
temperature.

The Euclidean action $S_F (A,M)$
is given by
\be
S_F (A,M) \;=\; \int_0^\beta d \tau \int d^2 x \; {\bar \psi}
( \not \! \partial + i e \not \!\! A + M ) \psi \; .
\ee
We are using Euclidean Dirac's matrices in the representation
\be
\gamma_1 = \sigma_1 \;\;\;\;\gamma_2 = \sigma_2 \;\;\;\;\gamma_3 = \sigma_3
\ee
where $\sigma_i$ are the familiar Pauli matrices
and $\beta = {1}/{T}$ is the inverse temperature. The label
$3$ is used to denote the Euclidean time coordinate $\tau$.
The fermionic fields in (\ref{ealagamma}) obey  antiperiodic boundary
conditions in the timelike direction
\be
\psi (\beta , x) \;=\; - \, \psi (0 , x) \;\;\;\;,\;\;\;\;
{\bar \psi} (\beta,x) \;=\; - {\bar \psi} (0, x) \;\;, \forall x \;,
\label{fbc}
\ee
with $x$ denoting the two space coordinates. The gauge field
satisfies periodic boundary conditions instead
\be
A_\mu (\beta,x) \;=\; A_\mu (0,x) \;\;,\;\; \forall x \;.
\label{gbc}
\ee

We want to make a
calculation which preserves an interesting property of the imaginary time
formulation, namely, that there is room for gauge transformations with
non-trivial winding around the time coordinate, and any approximation
which assumes the smallness of $A_3$ may put the symmetry under
those large transformations in jeopardy.

Let us first discuss the non-trivial gauge transformations at
finite temperature.
The set of allowed gauge transformations in the
imaginary time formalism is defined in the usual way:
$$\psi (\tau,x) \; \to \; e^{-i e \Omega (\tau,x)} \psi (\tau,x) \;\;,\;\;
{\bar \psi} (\tau,x) \; \to \; e^{i e \Omega (\tau,x)} {\bar \psi}
(\tau,x)$$
\be
A_\mu (\tau,x) \;\to \; A_\mu (\tau,x) \,+\, \partial_\mu \Omega (\tau,x)
\ee
where $\Omega(\tau,x)$ is a differentiable function vanishing at spatial
infinity
($|x| \to \infty$), and whose time boundary conditions
are chosen in order not to affect the fields' boundary conditions
(\ref{fbc}) and (\ref{gbc}). It turns
out that $\Omega(\tau,x)$ can wind an arbitrary number of times around
the cyclic time dimension:
\be
\Omega(\beta,x) \;=\; \Omega(0,x) \,+\, \frac{2 \pi}{e} \, k
\label{omegabc}
\ee
where $k$ is an integer which labels the homotopy class of the gauge
transformation.

Invoking gauge invariance of the fermionic
determinant
\be
\det ( \not \! \partial + i e \not \!\! A \,+\, M ) \; = \; \int {\cal D}
\psi \, {\cal D} {\bar \psi}
\; \exp \left\{ - \int_0^\beta d \tau \int d^2 x {\bar \psi} ( \not \!
\partial + i e \not \! A \,+\, M )
\psi \right \} \;,
\label{fdet}
\ee
we can always perform a gauge transformation of the fermionic fields in the
functional
integral (\ref{fdet})
\be
\psi (\tau,x) \;=\; e^{-i e \Omega (\tau,x)} \psi' (\tau,x) \;\;\;\;
{\bar \psi} (\tau,x) \;=\; e^{i e \Omega (\tau,x)} {\bar \psi}' (\tau,x)
\label{fchange}
\ee
in order to pass to an equivalent expression where the gauge field is traded
for $A_\mu' = A_\mu + \partial_\mu \Omega$:
\be
\det ( \not \! \partial + i e \not \! A \,+\, M ) \; = \; \int {\cal D}
\psi' \,
{\cal D} {\bar \psi}'\; \exp \left\{ - \int_0^\beta d \tau \int d^2 x {\bar
\psi}'
( \not \! \partial + i e \not \! A' \,+\, M )\psi' \right \} \;,
\label{fdett}
\ee

We consider the configurations given by eqs.(\ref{0}) and (\ref{1}),
namely $A_3$ is only a function of $\tau$, and $A_j$ is
independent of $\tau$. Under these assumptions, we see that the only
$\tau$-dependence of the Dirac operator comes from $A_3$. This
dependence can however be erased by a redefinition of the integrated
fermionic fields like in eq.(\ref{fchange}) if we take
\be
\Omega (\tau) \;=\;
- \int_0^\tau d \tau ' A_3 (\tau ') +
\left( \frac{1}{\beta} \int_0^\beta d \tau ' A_3 (\tau ')
+\frac{2\pi k}{e\beta} \right) \tau
\ee
where $k$ is the arbitrary integer labeling the homotopy class.
Such a transformation renders $A_3'$ constant.
The freedom to choose $k$ could be used to further restrict the
values of the constant $A_3'$
\be
 0 \leq   A_3' < \frac{2 \pi}{e\beta}
\ee
or any of the intervals obtained by a translation of this one by an integer
number of $\frac{2 \pi}{e \beta}$. In this sense, the value of the constant
in such an interval is the only `essential', i.e., gauge invariant,
$A_3 (\tau)$ dependent information
contained in the configurations (\ref{0})-(\ref{1}), describing the holonomy
$\int_0^\beta d {\tilde \tau} A_3 ({\tilde \tau})$ around the time direction
(notice that the $F_{3j}$
components of the field curvature tensor identically vanish for this
configurations). However, we will limit ourselves to small gauge
transformations ($k=0$) in order to avoid any assumption about large gauge
invariance of the fermionic measure in (\ref{fdet}) and safely discuss the
effect of large gauge transformations on the final results. Thus the constant
field $A_3'$ takes the mean value of $A_3(\tau)$,
\be
{\tilde A}_3 \;=\; \frac{1}{\beta} \, \int_0^\beta \, d \tau \, A_3 (\tau)
\;.
\ee
Note that the spatial components of $A_\mu$ remain $\tau$-independent
after this redefinition.

After redefining the fermionic fields according to this prescription, we
see that the fermionic determinant we should consider is now
\be
\det ( \not \! \partial + i e \not \! A \,+\, M ) \; = \; \int {\cal D}
\psi \,
{\cal D} {\bar \psi}\; \exp [- S_F ( A_j , {\tilde A}_3 , M)  ] \;,
\label{fdet1}
\ee
where
\be
S_F ( A_j , {\tilde A}_3 , M) \;=\; \int_0^\beta d \tau \int d^2 x \; {\bar
\psi}
( \not \! \partial + i e ( \gamma_j A_j + \gamma_3 {\tilde A}_3 ) + M )
\psi \; ,
\ee
and we removed the primes for the sake of clarity.

Since the Dirac operator
in the previous equation is invariant under imaginary time translations
it is convenient to perform a Fourier
transformation on the time variable for $\psi$ and ${\bar \psi}$
\ba
\psi (\tau, x) &=& \frac{1}{\beta} \, \sum_{n=-\infty}^{+\infty} \,
e^{i \omega_n \tau} \psi_n (x) \nonumber\\
{\bar \psi} (\tau, x) &=& \frac{1}{\beta} \, \sum_{n=-\infty}^{+\infty} \,
e^{-i \omega_n \tau} {\bar \psi}_n (x) \;,
\label{Four}
\ea
where $\omega_n = (2 n +1) \frac{\pi}{\beta}$ is the usual Matsubara
frequency for fermions.
Then the Euclidean action is written as an infinite series of decoupled
actions, one for each Matsubara mode
\be
S_F ( A_j , {\tilde A}_3 , M) \;=\; \frac{1}{\beta}
\sum_{n=-\infty}^{+\infty}
\int d^2 x {\bar \psi}_n (x) \left[ \not \! d \,+\, M \,+\, i \gamma_3
(\omega_n + e {\tilde A}_3) \right] \psi_n (x)
\ee
where $\not \!\! d$ is the $1+1$ Euclidean Dirac operator
corresponding to the spatial
coordinates and the spatial components of the gauge field
\be
\not \! d \;=\;\gamma_j (\partial_j + i e A_j).
\label{sup}
\ee
As the action splits up into a series and the fermionic measure
can be written as
\be
{\cal D} \psi(\tau,x) \, {\cal D} {\bar \psi}(\tau,x)=
\prod_{n=-\infty}^{n=+\infty}{\cal D} \psi_n(x) \, {\cal D} {\bar \psi}_n(x)
\label{measure1}
\ee
the $2+1$ determinant is an infinite product of the corresponding $1+1$
Euclidean Dirac operators
\be
\det ( \not \! \partial + i e \not \! A \,+\, M ) \; = \;
\prod_{n=-\infty}^{n=+\infty} \det [\not \! d + M + i \gamma_3 (\omega_n
+ e {\tilde A}_3) ] \;.
\ee
Explicitly, the $1+1$ determinant for a given mode is a functional integral
over $1+1$ fermions
\begin{eqnarray}
\lefteqn{\det [\not \! d + M + i \gamma_3 (\omega_n
+ e {\tilde A}_3) ]  = } \nonumber \\
& &  \int {\cal D} \chi_n \, {\cal D} {\bar \chi}_n \;
\exp\left\{ - \int d^2 x {\bar \chi}_n (x) ( \not \! d + M + i \gamma_3
(\omega_n + e {\tilde A}_3 ) ) \chi_n (x) \right\} \;.
\label{fori1}
\end{eqnarray}

In order to compute $\Gamma_{odd}$ we factorize now these determinants in a
piece which is sensitive to the sign of $M$ and a piece which is not.
The Euclidean action $S_n$ corresponding to the mode $n$ may be
conveniently recasted in the following form
\be
S_n \;=\; \int d^2 x \; {\bar \chi}_n ( \not \! d +
\rho_n e^{i \gamma_3 \phi_n} ) \chi_n
\ee
with
\be
\rho_n \;=\; \sqrt{ M^2 + ( \omega_n + e {\tilde A}_3 )^2 }\;;
\phi_n \;=\; {\rm arctan} ( \frac{\omega_n + e {\tilde A}_3}{M} ) \;.
\ee
We next realize that the change of fermionic variables
\be
\chi_n (x) \;=\; e^{- i \frac{\phi_n}{2} \gamma_3} {\chi'}_n (x) \;\;,\;\;
{\bar \chi}_n (x) \;=\; {{\bar \chi}'}_n (x) e^{- i \frac{\phi_n}{2}
\gamma_3} 
\label{quiral}
\ee
makes the action $S_n$ independent of $\phi_n$.
This is not a gauge transformation but a global chiral rotation in the
$1+1$ Euclidean fermionic variables. Correspondingly,
the fermionic measure picks up an
anomalous Fujikawa jacobian \cite{fuji} so that one ends with
\be
\det [\not \! d + M + i \gamma_3 (\omega_n
+ e {\tilde A}_3) ] \;=\; J_n[A,M] \; \det [\not \! d +  \rho_n ]
\label{fuji1}
\ee
where
\be
J_n[A,M] \;=\; \exp ( -i \frac{e \phi_n}{2 \pi} \int d^2 x \epsilon_{jk}
\partial_j A_k )\;,
\label{J}
\ee
with $\epsilon_{jk}$ denoting the $1+1$ Euclidean Levi-Civita symbol.

Recalling the definition of $\Gamma_{odd}$, we see that the second factor
in expression (\ref{fuji1}) does not contribute to it, since it is
invariant under $M \to -M$. The Jacobian (\ref{J}), instead, changes to
its inverse. As a consequence, the parity odd
piece in the effective action is given in terms of the infinite set
of $n$-dependent Jacobians:
\be
\exp [ - \Gamma_{odd} ] \;=\; \prod_{n=-\infty}^{n=+\infty} \; J_n[A,M]
\ee
or
\be
\Gamma_{odd} \;=\; - \sum_{n=-\infty}^{n=+\infty} \, \log J_n[A,M]
         \;=\; i \frac{e}{2 \pi} \, \sum_{n=-\infty}^{n=+\infty}
\phi_n \; \int d^2 x \epsilon_{jk} \partial_j A_k \;.
\ee
There only remains to perform the summation over the $\phi_n$'s.
This can be done by using standard techniques in Finite Temperature Field
Theory. We define
\be
{\cal S} \;=\;\sum_{n=-\infty}^{n=+\infty} \, {\rm arctan}
( \frac{\omega_n + e {\tilde A}_3}{M} )\,,
\ee
whose sign will obviously depend on the sign of $M$. We make
this explicit by rewriting ${\cal S}$ as
\be
{\cal S} \;=\; \frac{M}{|M|} \, \sum_{n=-\infty}^{n=+\infty} \, {\rm arctan}
( \frac{\omega_n + e {\tilde A}_3}{|M|} )
%\label
\ee
or, using the expression for $\omega_n$
\be
{\cal S}(x,y) \;=\; \frac{M}{|M|}
\sum_{n=-\infty}^{n=+\infty} \, {\rm arctan}
( \frac{(2 n + 1) \pi + x}{y} )
\ee
where $x = e \beta {\tilde A}_3$, and $y = \beta |M|$ are the two
dimensionless parameters built from the original ones.
This series must be regularized, and the standard technique consists in
subtracting the zero-field value of each term; notice that the sum of
these zero-field contributions conditionally converges to 0.
Then
\be
{\cal S}(x,y)
\;=\; \frac{M}{|M|}
\sum_{n=-\infty}^{n=+\infty} \,\int_0^x  du
\frac{d}{du} {\rm arctan}( \frac{(2 n + 1) \pi + u}{y} )
\label{intyder}
\ee
As the series now converges absolutely we can first perform the summation.
The sum to be evaluated is then
\be
\sum_{n=-\infty}^{n=+\infty} \, \frac{y}{y^2 + [ (2 n + 1) \pi + u ]^2} \;
\ee
which is solved by the summation formula
\be
\sum_{n=-\infty}^{n=+\infty} \, \frac{1}{(n-x_1)(n-x_2)} \;=\;
- \frac{\pi({\rm cot}(\pi x_1) - {\rm cot}(\pi x_2)}{x_1-x_2}.
\label{2.96}
\ee
After performing the integral we get
\be
{\cal S} \;=\;  \frac{M}{|M|}
{\rm arctan} \left[
\tanh(\frac{\beta |M|}{2}) \tan ( \oh e \beta {\tilde A}_3 ) \right] .
\label{suma1}
\ee
Thus the parity-odd part of $\Gamma$
finally reads
\be
\Gamma_{odd} \;=\; i \frac{e}{2\pi} \frac{M}{|M|}
{\rm arctan} \left[ \tanh(\frac{\beta |M|}{2})
\tan ( \frac{e}{2} \int_0^\beta d \tau
A_3(\tau) ) \right]\, \int d^2 x \epsilon_{jk} \partial_j A_k \; .
\label{espl'}
\ee

There are several observations to be made about our result (\ref{espl'}).
First
we observe that this result has the proper
zero temperature limit
\be
\lim_{T \to 0} \Gamma_{odd} \ \; \to \;\frac{1}{2}  \frac{M}{|M|}S_{CS}
\label{T=0}
\ee
where $S_{CS}$ is the Chern-Simons action
 \be
S_{CS}=i \frac{e^2}{4\pi}\int d^3x \epsilon_{\mu\nu\alpha}A_\mu
\partial_\nu A_\alpha
\label{SCS}
\ee
which shows up in our particular configuration (\ref{0})-(\ref{1}) as
$\frac{e^2}{2\pi} \int  d\tau A_3(\tau)\int d^2x \epsilon_{ij}\partial_i A_j$.
So we get the
induced Chern-Simons term at zero temperature.
As it is well known, in the zero temperature case the result is not
invariant under large gauge transformations. The quantization of
the spatial integral that measures the flux of the magnetic field through
a space-like manifold $\tau=constant$ in units
of $\frac{2\pi}{e}$ shows that (\ref{T=0}) changes 
by the addition of an odd multiple of $i \pi$
under a large gauge transformation with odd winding number when the 
magnetic flux is odd. This gauge non-invariance is compensated by the parity 
anomaly discussed in the Introduction when the complete result is
regularized in a gauge invariant scheme.

The same situation occurs in the finite temperature result (\ref{espl'}).
A large gauge transformation with odd winding number $k=2p+1$ shifts the 
argument of the tangent in $(2p+1)\pi$. Although the tangent is not sensitive 
to such a change, one has to keep track of it by shifting the branch used for 
arctan definition. This amounts to the same result as in the $T \to 0$ limit:
the gauge non-invariance of $\Gamma_{odd}$ under large gauge transformations 
is compensated by the parity anomaly $\pm \frac{1}{2} S_{CS}$.

Now we observe that a perturbative expansion in terms of $e$ yields the usual
perturbative result
\be
\Gamma_{odd} = \frac{1}{2}  \frac{M}{|M|} {\tanh}(\frac{|M|\beta}{2})S_{CS}
+O(e^4)
\ee
where the coefficient of the Chern-Simons term acquires a smooth
dependence on the temperature.
Were we considering only the first non trivial order in perturbation theory,
we would find a clash between temperature dependence and gauge invariance
\cite{bfs}-\cite{cfrs}: the gauge non-invariance of the induced CS term is no 
longer compensated by the parity anomaly.
Now we learn, as it was stressed in \cite{dll}
in a $0+1-$dimensional example and in \cite{dgs} in $2+1$ dimensions,
that one has to consider the full result in order to analyse gauge
invariance.

Finally, we observe that the result (\ref{espl'}) is
not an extensive quantity
in Euclidean time. It is however extensive in space, and that is indeed
all one expects in Finite Temperature Field Theory. In contrast, the $T=0$
limit becomes an extensive quantity in space-time, as is expected from zero
temperature Field Theory.

We shall now extend the previous results, obtained for
space-independent $A_3$ and time-independent $A_j$ to the somewhat
more general situation of a smooth spatial dependence
of $A_3$  besides the previous arbitrary time dependence.

The fermionic determinant we should calculate, after getting rid of the
$\tau$ dependence of $A_3$ will have a form analogous to (\ref{fdet1})
with the only difference of having an $x$ dependence in ${\tilde A}_3$.
As there is no explicit time-dependence in the Dirac
operator, we again pass to a Fourier
description of the time coordinate.
Defining the $x$-dependent fields $\rho_n (x)$ and $\phi_n (x)$,
\be
\rho_n (x)\;=\; \sqrt{ M^2 + ( \omega_n + e {\tilde A}_3 (x) )^2 }\;; ~~~~~~~
\phi_n (x)\;=\; {\rm arctan} ( \frac{\omega_n + e {\tilde A}_3 (x)}{M} )\;,
\ee
we have for the complete fermionic determinant an expression equivalent to
the previous case:
\be
\det ( \not \! \partial + i e \not \! A \,+\, M ) \;=\;
\prod_{n=-\infty}^{\infty} \, \det \left[ \not \!\! d + \rho_n (x)
e^{i \gamma_3 \phi_n (x)} \right] \;.
\ee
The determinant corresponding to the $n$-mode is again written as
a functional integral over $1+1$ dimensional fields, but a
transformation like (\ref{quiral}) is now a 
{\em local} chiral rotation of the $1+1$ dimensional fermions
and gives rise to
\be
\det \left[ \not \!\! d + \rho_n (x)
e^{i \gamma_3 \phi_n (x)} \right] \;=\;
J_n \; \det [\not \! d' +  \rho_n (x)] \;,
\label{fact}
\ee
where:
\be
\not \! d' \;=\; \not \! d \,-\, \frac{i}{2} \not \! \partial \phi_n
\gamma_3 \;
\ee
and the anomalous Jacobian reads
\be
J_n \;=\; \exp \left\{-i \frac{e }{2 \pi} \int d^2 x [ \phi_n (x)
\epsilon_{jk}
\partial_j A_k + \frac{1}{4} \phi_n (x) \Delta \phi_n (x) ] \right\} \;.
\ee

The $x$-dependence of the phase factor $\phi_n$ affects the result in two
ways: first, we see that the field redefinition changes the operator
$\not \! d $ to $\not \! d'$ which depends on the sign of $M$, and so
there will be a contribution to $\Gamma_{odd}$ coming from the determinant
of $\not \! d' +  \rho_n (x)$. Second, the Jacobian is now a more involved
function of $\phi_n$, since the field redefinition affects the Dirac
operator
which is used to define the fermionic integration measure. In a first
approximation,
we shall only take into account the contribution coming from the Jacobian,
since the one that follows from the determinant of the Dirac operator is of
higher order in a derivative expansion (and we are assuming that the
$x$-dependence
of ${\tilde A}_3$ is smooth).
The contribution which is quadratic in $\phi_n$ is irrelevant to the
parity breaking piece, since it is even in $M$.
Thus, neglecting the terms containing
derivatives of ${\tilde A}_3$, we
have for $\Gamma_{odd}$ a result which looks like a natural generalization
of
the previous case
\be
\Gamma_{odd} \;=\; i \frac{e}{ 2\pi} \frac{M}{|M|}  \int d^2 x
\; {\rm arctan} \left[ \tanh(\frac{|M|\beta}{2})
\tan ( \frac{e}{2} \int_0^\beta d \tau
A_3(\tau, x) ) \right]\,  \epsilon_{jk} \partial_j A_k (x)\;.
\ee
It is not hard to check that the reliability of the approximation of
neglecting derivatives of ${\tilde A_3}$ is assured if the condition
\be
| e \; \partial_j {\tilde A_3} | \;<<\; M^2
\ee
is fulfilled. To end with this example, let us point that all
the remarks we made for the case of a space-independent $A_3$
also apply to this case.
%%%%%%%%%%%%%%%%%%%%%%%%%%%%%%%%%%%%%%%%%%%%%%%%%%%%%%%%%%%%%%%%%%%%
%%%%%%%%%%%%%%%%%%%%%%%%%%%%%%%%%%%%%%%%%%%%%%%%%%%%%%%%%%%%%%%%%%%%
%%%%%%%%%%%%%%%%%%%%%%%%%%%%%%%%%%%%%%%%%%%%%%%%%%%%%%%%%%%%%%%%%%%%
%%%%%%%%%%%%%%%%%%%%%%%%%%%%%%%%%%%%%%%%%%%%%%%%%%%%%%%%%%%%%%%%%%%%
\section{The non-Abelian case}
We extend in this section the previous analysis to 
the non-Abelian case.
Although we shall
consider, as in the Abelian case,
particular background field configurations which
allow to make exact
computations, the results will exhibit genuine
non-Abelian effects through
non trivial commutators of spatial components of the gauge field.
Our analysis will be valid for the $SU(N)$ case
although  some points are made explicit for the particular $N=2,3$
cases. As we shall see, details arising in calculations 
are due to technicalities associated with handling the
non-Abelian symmetry; once they are overcome, the results
appear as a natural extension of the Abelian ones.

The Euclidean fermionic action which describes the system is now written as
\be
S_F (A,M) \;=\; \int_0^\beta d \tau \int d^2 x \; {\bar \psi}
( \not \!\! D  + M ) \psi \;,
\label{nabact}
\ee
where the covariant derivative acting on the fermions in the fundamental
representation of $SU(N)$ is defined as 
\be
D_\mu = \partial_\mu \,+
ig \, A_\mu \; ,
\label{ss}
\ee 
and the gauge connection $A_\mu$ is written as
\be
A_\mu \;=\;  A_\mu^a \, \tau_a
\ee
with $\tau_a$ denoting hermitian generators of the Lie algebra
($a=1, \dots, N^2-1$), verifying the relations
\be
\tau_a^\dagger \;=\;  \tau_a ,\;\;\;
[\tau_a , \tau_b] \;=\;  i f_{a b c} \tau_c ,\;\;\;
{\rm tr} (\tau_a \tau_b) = \frac{1}{2} \delta_{a b} \;,          %(56)
\ee
with $f_{a b c}$  the totally antisymmetric
structure constants. For the particular
case of $SU(2)$, which we shall consider in more detail, we have
$f_{a b c}=\epsilon_{a b c}$ since the generators 
will be taken to
be the usual Pauli matrices.

We are concerned with the parity-odd piece of the efective action defined in
(\ref{gammaodd}). Fermionic (bosonic) fields satisfy again 
antiperiodic (periodic) 
boundary conditions in the timelike direction.

We shall in this case restrict the set of configurations for the
gauge fields given by (\ref{0})-(\ref{aj}) in order to be able to 
calculate $\Gamma_{odd}$ exactly.
Before doing so, let us clarify a point about the
nature of the gauge group boundary conditions in imaginary time.

Non-Abelian gauge transformations are defined by their action
on the fermionic and gauge fields
$$
\psi (\tau,x) \;\to\; \psi^U (\tau,x) \;=\; U(\tau,x) \, \psi (\tau,x), 
\;\;\;\;\;\;
{\bar \psi} (\tau,x) \;\to\; {\bar \psi}^U (\tau,x) \;=\;
\psi (\tau,x)\,U^\dagger (\tau,x)
$$
\be
A_\mu (\tau,x) \;\to \; A_\mu^U (\tau,x) \;=\;
U(\tau,x) A_\mu (\tau,x) U^\dagger (\tau,x)
\,-\frac{i}{g}\, U(\tau,x) \partial_\mu U^\dagger (\tau,x) \;.
\ee
In order to decide the boundary conditions the gauge group element
should satisfy in the timelike direction, one requires that the
periodicity of the gauge field and the antiperiodicity
of the fermions is unaltered under a gauge transformation. 
Concerning the gauge field,
this only imposes on $U$ the condition
\be
U(\beta,x) \;=\; h \, U(0,x)
\label{cnter}
\ee
where $h$ is an element of $Z_N$, the center of $SU(N)$.
Now, concerning  fermions, the condition on $U$  depends
on whether they are in the fundamental or adjoint representation.
In the fundamental one, it is easily seen that
\be
U(\beta,x) \;=\; U(0,x)
\ee
while in the adjoint representation, condition (\ref{cnter}) follows
instead. As we assume fermions are in the fundamental representation,
the group elements $U(\tau,x)$ are taken to be strictly periodic (a
condition in fact analogous to the one used for the Abelian case in
eq.(\ref{omegabc})). One can then prove \cite{alvarez} that for compact 
groups
\be
w(U)=\frac{1}{12 \pi^2 N}tr \int d^3x \epsilon_{\mu\nu\alpha}
U^{-1}\partial_{\mu}U U^{-1}\partial_{\nu}U U^{-1}\partial_{\alpha}U
\label{winding}
\ee
is an integer number that labels homotopically equivalent gauge 
transformations. Thus the disctintion between large and small gauge 
transformations has a different origin here than in the Abelian case.

We thus consider a class of configurations equivalent by gauge 
transformations to
\be
A_3 = |A_3|(\tau) \check{n}, 
\label{rset0}
\ee
\be
 A_j = A_j(x) \,,\,\, 
[A_j , \check{n}]=0
\,\,\, (j=1,2) \;.
\label{rset1}
\ee
where $\check{n}$ is a fixed direction in the Lie algebra
($\check{n}=n^a\tau_a$, $ n^an^a=1$).

We note that conditions (\ref{rset0})-(\ref{rset1}) 
assure the vanishing of the colour
electric fields, as well as the time independence of the
colour magnetic fields. Regarding the condition
(\ref{rset1}), which requires the spatial gauge field
components to commute with $A_3$, it is worth remarking that
its consequences depend strongly  on whether the group considered
is $SU(2)$ or $SU(N)$ with $N>2$. In the former case, the only
solution to  (\ref{rset1}) corresponds
to a configuration with all the
gauge field components pointing in the same direction $\check{n}$ in internal
space, i.e.  an  `Abelian like' configuration.
In contrast, for $N>2$, configurations with $[A_1,A_2] \neq 0$
are indeed possible. 

To make the point above more explicit let us analyse 
the simple specific example of $SU(3)$
with the generators given by the standard Gell-Mann matrices; 
one can then take
$A_1$ and $A_2$ as linear combinations of $\tau_1$, $\tau_2$ and $\tau_3$
(generators of a $SU(2)$ subgroup) and $A_3$ pointing in the direction of 
$\tau_8$. This situation easily generalizes to $N>3$
since one can construct the set of
generators for a higher $N$ in such a way that it contains the
generators corresponding to $SU(N-1)$ as a subset of block-diagonal
matrices, and one of the extra generators can be always
defined as to commute with them.
Thus it is possible to take $A_1$ and 
$A_2$ as non commuting vectors in the subalgebra corresponding to 
$SU(N-1)$ and $A_3$ commuting with them. 

Coming back to the general case, let us point that,
as in the Abelian case, one can erase the $\tau$ dependence of $A_3$ 
component by considering a change of variables for  the fermionic fields
corresponding to a gauge transformation of the form
\be 
U(t)=e^{ig\Omega(\tau)\check{n}}
\label{borratau}
\ee
and
\be
\Omega (\tau) \;=\;
- \int_0^\tau d { \tau '} A_3^{(\check{n})} ({ \tau '}) +
\left( \frac{1}{\beta} \int_0^\beta d { \tau '} A_3^{(\check{n})}
({\tau '}) \right) \tau.
\ee
Now, because of  condition (\ref{rset1}) the space components of
the gauge field remain unchanged under this transformation, while 
$A_3$ takes the constant value $\tilde{A}_3 =
\frac{1}{\beta}\int_0^{\beta} d\tau A_3(\tau) = |\tilde{A_3}|\check{n}$.
After these remarks, we assume a gauge transformation has been made on
the fermions in order to reach a constant $\tilde{A}_3$ and the rest of 
conditions (\ref{rset0})-(\ref{rset1}) for the gauge field.

After a  Fourier transformation on the time
variable for $\psi$ and ${\bar \psi}$ of the form (\ref{Four})
the Euclidean action
can be written as an infinite series of
decoupled actions,
\be
S_F \;=\;
\frac{1}{\beta}
\sum_{n=-\infty}^{+\infty}
\int d^2 x {\bar \psi}_n (x) \left[ \not \! d \,+\, M \,+\,i \gamma_3
(\omega_n + g \tilde{A}_3^a \tau_a) \right] \psi_n (x)
\label{Four2}
\ee
where $\not \!\! d=\;\gamma_j (\partial_j +i g A_j)$ is the non-Abelian
Dirac operator corresponding to the spatial coordinates and the
spatial components of the gauge field. Concerning
the fermionic measure, we write it in the form
\be
{\cal D} \psi(\tau,x) \, {\cal D} {\bar \psi}(\tau,x)=
\prod_{n=-\infty}^{n=+\infty}{\cal D} \psi_n(x) \, {\cal D}
{\bar \psi}_n(x) \;,
\label{measure}
\ee
so that again the $2+1$ determinant becomes an infinite product of the
corresponding $1+1$ Euclidean Dirac operators
\be
\det ( \not \! \partial +ig \not \! A \,+\, M ) \; = \;
\prod_{n=-\infty}^{n=+\infty}
\det [\not \! d + M + i \gamma_3 (\omega_n +g \tilde{A}_3^a \tau_a) ] \;.
\label{xx}
\ee
We now show that the same trick which lead to the decoupling
of parity breaking and parity conserving parts of the determinant
for the Abelian case can be applied here.
First, we use the property
\be
M + i \gamma_3 (\omega_n +g \tilde{A}_3^a \tau_a) \;=\;
\rho_n \, e^{i \phi_n } 
\label{trick1}
\ee
where
\be
\rho_n \;=\; \sqrt{ M^2 + ( \omega_n + g \tilde{A}_3^a \tau_a )^2 }\;;
\phi_n \;=\; {\rm arctan} ( \frac{\omega_n + g \tilde{A}_3^a \tau_a}{M} ) \;.
\ee

The usual definition of functions of matrices in terms of power
series has been used above. It is important to realize that,
being $\phi_n$ a non-trivial Hermitean function of a matrix in the Lie
algebra, it will in general have components along the generators $\tau_a$
and also along the identity
matrix, namely,
\be
\phi_n \;=\; \phi_n^0 1 + \phi_n^a \tau_a \;.
\ee
As an illustration, we consider the $SU(2)$ case. A somewhat lenghty but 
otherwise straightforward calculation
yields explicit expressions for these components of $\phi_n$:
\ba
\phi_n^0 &=& \frac{1}{2} \, \arctan \left(
\frac{2 M \omega_n}{M^2 + \frac{g^2}{4} |\tilde{A}_3|^2 - \omega_n^2} \right)
\nonumber\\
\phi_n^a &=& \arctan \left(
\frac{g M |\tilde{A}_3|}{M^2 - \frac{g^2}{4} |\tilde{A}_3|^2 + 
\omega_n^2} \right) n^a \;.
\label{phis}
\ea

The $1+1$ determinant for a given mode is a functional
integral over $1+1$ fermions that using (\ref{trick1}) can be written as
\be
\det [\not \! d + M + i \gamma_3 (\omega_n +g \tilde{A}_3^a \tau_a) ] =
\int {\cal D} \chi_n \, {\cal D} {\bar \chi}_n \;
\exp\left\{ - \int d^2 x {\bar \chi}_n (x)
( \not \! d + \rho_n e^{i \gamma_3 \phi_n} )
\chi_n (x) \right\} \;.
\label{fori}
\ee

We now perform the change of fermionic variables
\be
\chi_n (x) \;=\; e^{- i \frac{\phi_n}{2} \gamma_3} {\chi'}_n (x) \;\;,\;\;
{\bar \chi}_n (x) \;=\; {{\bar \chi}'}_n (x) e^{- i \frac{\phi_n}{2}
\gamma_3} \;,
\label{chiral}
\ee
and verify that due to the last condition in (\ref{rset1}) it indeed 
decouples the parity violating part of the
effective action. We find, including the
anomalous Fujikawa Jacobian
\be
\det [\not \! d + M + ig \gamma_3 (\omega_n
+  \tilde{A}_3^a \tau_a) ] \;=\; J_n \; \det [\not \! d +  \rho_n ].
\label{fuji}
\ee
 
The Jacobian in (\ref{fuji}) reads \cite{fuji}
\be
J_n[A,M] \;=\; \exp \left[ -i tr \frac{\phi_n}{2}   
\int d^2x {\cal A} \right]\;,
\label{jaco}
\ee
with ${\cal A} =  {\cal A}^a\tau^a $
denoting the $1+1$ Euclidean anomaly under
an infinitesimal non-Abelian axial transformation. As this
transformation is $x$-independent, there is no difference between
finite and infinitesimal transformations and one
can just simply iterate the infinitesimal Fujikawa
Jacobian \cite{fuji} in order to get the finite answer
(\ref{jaco}). Also note that
$\phi_n^0$ (the component along the
identity) does not contribute to the jacobian since
$tr (\phi_n^0 {\cal A}) = 0$. A standard calculation
leads for the two-dimensional
non-abelian anomaly the answer (see for example \cite{gss})
\be
{\cal A} = \frac{g}{2\pi} \epsilon_{ij} F_{ij}
\label{ano}
\ee
so that the Jacobian finally takes the form
\be
J_n[A,M] \;=\; \exp \left[ -  \frac{ig}{4\pi}
tr \left(
\phi_n  \int d^2x \, \epsilon_{ij} F_{ij} \right) \right] \; .
\label{jaco2}
\ee

We see from eqs.(\ref{xx}) and (\ref{fuji}) that
the parity odd piece of the effective action is again given in terms of
the infinite set of $n$-dependent Jacobians,
\be
\Gamma_{odd}[A,M] \;=\; - \sum_{n=-\infty}^{n=+\infty} \, \log J_n[A,M]
  \;=\; \frac{ig}{4\pi}
tr \left(  (\sum_{n=-\infty}^{+\infty} \phi_n )
\; \int d^2x \, \epsilon_{ij} F_{ij} \right).          %(78)
\label{sinsumar}
\ee

Now we have to perform the summation over the $\phi_n$'s.
A careful analysis of the steps performed in the Abelian case shows that
the result (\ref{suma1}) is valid for matrix valued gauge fields. 
%(a scalar parameter must
%be introduced to perform the integral and derivative with respect to $u$
%in eq.(\ref{intyder})). 
Thus we get
\be
\Gamma_{odd} \;=\;
\frac{ig}{4\pi} tr \left(\arctan[\tanh(\frac{\beta M}{2}) 
\tan(\frac{g}{2} \beta \tilde{A}_3)]
\int d^2x \, \epsilon_{ij} F_{ij}\right).
\label{for1}
\ee
This is the main result in this section, which extends eq. (\ref{espl'}) to 
$SU(N)$ background fields.

\vspace{5mm}

We can check this result by doing explicit computations with the components 
$\phi_n^a$ given in eq.(\ref{phis}) for the $SU(2)$ case. 
From eq. (\ref{sinsumar}),
\be
\Gamma_{odd}[A,M] \;=\; \frac{ig}{8\pi}
\sum_{n=-\infty}^{+\infty} \phi_n^a
\; \int d^2x \, \epsilon_{ij} F^a_{ij}  .         %(80)
\label{componentes}
\ee
Using eq. (\ref{phis}) we have to compute 
\be
\Sigma = 
\sum_{n=-\infty}^\infty
\arctan \left(\frac{g M |\tilde{A}_3|}{M^2 - \frac{g^2}{4}|{\tilde A}_3|^2
+ \omega_n^2} \right) \;,
\ee
or, in terms of dimensionless variables
\be
m = \beta M
\;\;\;\;\;\;
x = \frac{g}{2} \beta |\tilde{A}_3|
\label{vari}
\ee
\be
\Sigma(x,m) = \sum_{n=-\infty}^\infty
\arctan \left(\frac{2 m x}{m^2 - x^ 2
+ (2n + 1)^2 \pi ^2} \right) \;.
\ee
The sum is convergent, but in order to calculate $\Sigma$ it will 
be convenient to write
\be
\Sigma(x,m) =
\int_0^x du \frac{\partial \Sigma}{\partial u}(u,m).
\label{deri}
\ee
The implicit subtraction of a zero-field contribution vanishes term by term
in this case.

After some calculations, one has
\be
\frac{\partial \Sigma}{\partial x}(x,m) = 2m \sum_{n=-\infty}^\infty
\frac{m^2 + (2n+1)^2\pi^2 + x^2}{[m^2 + (2n+1)^2\pi^2 - x^2]^2 + 4m^2x^2}
\label{ds}                                     %(86)
\ee
One could now arrange this expression to use the summation formula 
(\ref{2.96}). With the purpose of illustration
we use instead the standard Regge-type trick
to rewrite (\ref{ds}) as a contour integral of the form
\be
\frac{\partial \Sigma}{\partial x}(x,m) =
-\frac{m}{2\pi i} \oint_C dz \tanh({z}/{2})
\frac{m^2 - z^2 + x^2}{[m^2 - z^2 - x^2]^2 + 4m^2x^2}
\label{con}
\ee
where $C$ is a contour including all the poles of $\tanh({z}/{2})$.
After continuing $C$ into the upper and lower half-planes to
pick up the $4$ poles of the fraction only, we end with
\be
\frac{\partial \Sigma}{\partial x}(x,m) =
\frac{i}{2} [\tanh(\frac{x - im}{2}) - \tanh(\frac{x + im}{2})]
\label{coni}
\ee
Using this expression in (\ref{deri}) we finally get
\be
\Sigma(x,m) = 2 \arctan [\tanh(m/2) \tan(x/2)]
\label{S}
\ee
so that $\Gamma_{odd}$ can be written as
\be
\Gamma_{odd} \;=\;
\frac{ig}{4\pi} \arctan[\tanh(\frac{\beta M}{2}) \tan(\frac{g}{4} \beta 
|\tilde{A}_3|)] n^a
\int d^2x \, \epsilon_{ij} F^a_{ij}.
\label{for12}
\ee
Finally, observing that $(n^a\tau_a)^{(2k+1)}=\frac{1}{2^{2k}}n^a\tau_a$ 
and only odd powers 
enter the expansions of the functions involved, we see that the result is 
identical to eq.(\ref{for1}).

\vspace{5mm}

In order to analyze the result (\ref{for1}) let us write it in the most 
general form
\be
\Gamma_{odd} \;=\;
\frac{ig}{4\pi} tr \left(\arctan[\tanh(\frac{\beta M}{2}) \tan(
\frac{g}{2}\int_0^{\beta}d\tau A_3(\tau) )]
\, \int d^2x \epsilon_{ij} F_{ij} \right)
\label{for122}
\ee
Then we note that
in the zero-temperature limit one has
\be
\lim_{T \to 0} \Gamma_{odd} \;=\; \frac{ig^2}{8\pi} \frac{M}{|M|} 
tr\left(
\int_0^\beta d\tau A_3(\tau)\int d^2x \, \epsilon_{ij} F_{ij}\right).
\label{una}
\ee
This result is the usual one, namely
\be
\lim_{T \to 0} \Gamma_{odd} \;=\;
\frac{1}{2} \frac{M}{|M|} S_{CS},
\label{lim}
\ee
restricted to the particular background we have considered.
Here $S_{CS}$ is the non-Abelian CS action
\be
S_{CS}= \frac{ig^2}{8\pi}\int d^3x
\epsilon_{\mu\nu\alpha} tr (F_{\mu\nu}A_{\alpha} -\frac{2}{3}A_{\mu}
A_{\nu} A_{\alpha})
\label{NACS}
\ee
which for a gauge field satisfying the restrictions (\ref{rset1}) reads
\be
S_{CS} = \frac{i g^2}{4\pi} tr \int d^3x \,A_3  \epsilon_{ij}
F_{ij}.
\label{SCS3}
\ee
We thus recover the zero-temperature result
first obtained in \cite{det} by calculating the v.e.v. of the
fermion current in a constant non-Abelian  field strength tensor background
or in \cite{NS0} in a static non-Abelian magnetic background like ours. 
We recall, however, that gauge invariance under large gauge transformations is
obtained only when the parity anomaly $\pm \frac{1}{2} S_{CS}$
is added to the mass- and temperature-dependent expression for $\Gamma_{odd}$.

We finally note that a perturbative expansion in powers of the coupling 
constant $g$ shows a smooth temperature dependence of the CS coefficient,
\be
\Gamma_{odd}=\frac{1}{2}\tanh(\frac{M\beta}{2})S_{CS}+O(e^4).
\label{pertna}
\ee
Concerning gauge invariance of the finite temperature result
we note that, in contrast to the Abelian case,
there is no room for large gauge transformations preserving the conditions
(\ref{rset0}) and (\ref{rset1}) under wich our result (\ref{for122}) was
obtained. We can only quote gauge invariance under small gauge transformations
that do not mix spatial and time components. However, we expect that the 
large gauge invariance apparently broken by the perturbative expansion 
(\ref{pertna}) should be recovered by the full result.

\section{Summary and discussion}  

We have been able to compute the exact form of the parity violating 
contribution to the finite temperature effective action for 2+1 
massive fermions in a restricted set of gauge 
backgrounds, both for Abelian and non-Abelian gauge groups.
Our computation reproduces the standard results both at zero temperature and/or
perturbation theory.

The Abelian case allows for a complete analysis of the gauge invariance under 
large transformations; we have found that the mass and temperature dependent
contribution is not invariant but its variation is cancelled (modulo $2\pi i$)
when the parity anomalous contribution $\pm \frac{1}{2} S_{CS}$ is 
incorporated. We recall that in the zero temperature limit the gauge invariant 
result contains two 
contributions in the form of CS terms, one arising canonically from the 
fermion mass parity violating term and the other coming from the necessary 
parity anomaly of the gauge invariant fermionic measure in odd dimensions.
The present analysis gives a closed answer to the puzzle of gauge invariance 
of the effective action at finite temperature: the perturbative result in which
the CS coefficient acquires a smooth dependence on the temperature is correct, 
but shows that any perturbative order is insufficient to maintain large gauge 
invariance.

The non-Abelian case follows the pattern described above in every detail. 
Although the restrictions imposed on the background fields do not allow 
the study 
of large gauge transformations, notice that the zero temperature
limit shows the presence of two CS contributions with appropriate coefficients
so as to cancel the gauge non invariance of each other. This strongly suggests 
that the same behavior is to be expected concerning large gauge transformations
at finite temperature.

\underline{Acknowledgements}: 
G.L.R. thanks Prof. R.Jackiw for discussions and 
kind hospitality at CTP, MIT. 
C.D.F. and G.L.R. are supported by CONICET,
Argentina. F.A.S. is partially  suported
by CICBA, Argentina and a Commission of the European Communities
contract No:C11*-CT93-0315. This work is
supported in part by funds provided by the U.S. Department of Energy
(D.O.E.)
under cooperative research agreement \# DF-FC02-94ER40818.


\begin{references}
\bibitem{djt} S.~Deser, R.~Jackiw and S.~Templeton, Phys. Rev. Lett.
  {\bf 48} (1982) 975; Ann. Phys. (N.Y.) {\bf 140} (1982) 372.
\bibitem{wi} E.~Witten Commun. Math. Phys {\bf 121} (1989) 351.
\bibitem{det} A.N.~Redlich, Phys. Rev. Lett. {\bf 52} (1984) 18;
  Phys. Rev. {\bf D29} (1984) 2366.
\bibitem{NS0} A.~J.~Niemi and G.~W.~Semenoff,
Phys. Rev. Lett. {\bf 51} (1983) 2077.
\bibitem{GRS} R.E.Gamboa Sarav\'{\i}, G.L.Rossini and F.A.Schaposnik,
Int. J. Mod. Phys. {\bf A11} (1996) 2643.
\bibitem{pis} R.~Pisarski, Phys. Rev. {\bf D35} (1987) 664.
\bibitem{NS} A.J. Niemi and G.W. Semenoff,
Phys. Rev. Lett. {\bf 51} (1983) 2077.
\bibitem{N} A.J. Niemi, Nucl. Phys. {\bf B251} (1985) 55.
\bibitem{NS1} A.J. Niemi and G.W. Semenoff,
Phys. Rep. {\bf 135} (1986) 99.
\bibitem{DP} K.~Babu, A.~Das and P.~Panigrahi, Phys. Rev. {\bf D36}
(1987) 3725.
\bibitem{DP1} A.~Das and S.~Panda, J.~Phys.~A: Math. Gen.
{\bf 25} (1992) L245.
\bibitem{AF} I.J.R. Aitchinson, C.D. Fosco and J. Zuk,
Phys. Rev. {\bf D48} (1993) 5895.
\bibitem{P} E.R. Poppitz, Phys. Lett. {\bf B252} (1990) 417.
\bibitem{Bu} M.Burgess, Phys. Rev. {\bf D44} (1991) 2552.
\bibitem{K} W.T. Kim, Y.J. Park, K.Y. Kim and Y. Kim,
Phys. Rev. {\bf D 46} (1993) 3674.
\bibitem{I} K. Ishikawa and T. Matsuyama, Nucl. Phys. {\bf B 280}
[F518] (1987) 523.
\bibitem{bfs} N.~Brali\'c, C.D.~Fosco and F.A.~Schaposnik, Phys. Lett.
{\bf B 383} (1996) 199.
\bibitem{cfrs} D.~Cabra, E.~Fradkin, G.L.~Rossini and F.A.~Schaposnik,
Phys. Lett. {B 383} (1996) 434.
\bibitem{dll} G.~Dunne, K.~Lee and Ch.~Lu, Phys. Rev. Lett.
{\bf 78} (1997)
3434
\bibitem{dgs} S.~Deser, L.~Griguolo and D.~Seminara,
{\it Gauge Invariance,
Finite Temperature and Parity Anomaly in $D=3$},
hep-th/9705052 report.
\bibitem{frs} C.D.~Fosco, G.L.~Rossini and F.A.~Schaposnik,
{\it Induced Parity Breaking Term at Finite Temperature},
Phys. Rev. Lett. (in press), hep-th/9705124 report.
\bibitem{fuji} K.~Fujikawa, Phys. Rev. Lett. {\bf 42} (1979) 1195;
Phys. Rev. {\bf D21} (1980) 2848.
\bibitem{alvarez} O.Alvarez, Commun. Math. Phys. {\bf 100} (1985) 279.
\bibitem{gss} R.E.Gamboa Sarav\'\i ~,
F.A.~Schaposnik and J.E.~Solomin, Nucl.Phys. {\bf B185} (1981) 239.

\end{references}
\end{document}